# Next-to-leading Corrections at Small $x$*

S. Catani[a] and F. Hautmann[b]

[a]INFN, Sezione di Firenze, Largo E. Fermi 2, I-50125 Florence, Italy

[b]Cavendish Laboratory, University of Cambridge, Madingley Road, Cambridge CB3 0HE, UK

We discuss the role of sub-leading corrections to deep inelastic processes in the small-$x$ regime, and report recent results on the calculation of coefficient functions and quark anomalous dimensions.

## 1. Deep inelastic processes at small $x$

A new kinematic regime is explored at present (and future) hadron colliders. It is characterized by small values of the ratio $x = Q^2/S$ between the typical momentum $\sqrt{Q^2}$ transferred in the hard scattering and the total centre-of-mass energy $\sqrt{S}$. The most striking phenomenological feature at small $x$ is the strong rise of the deep inelastic cross sections. This has been recently observed by the HERA collaborations [1], who have measured the structure function $F_2$ down to $x$-values as low as $10^{-3} \div 10^{-4}$. Qualitatively, such a behaviour can be accounted for within QCD by a leading-logarithmic approximation in either $\ln Q^2$ or $\ln x$ [2]. However, precise quantitative comparisons of the theory with experiment require QCD corrections to be controlled in sub-leading orders.

The standard analysis of the $\ln Q^2$-dependence in deep inelastic scattering is based on the renormalization group evolution equations [3], and on the expansion of anomalous dimensions and Wilson coefficient functions to fixed order in $\alpha_S$. These quantities are at present fully known up to two-loop order [4,5].

Nonetheless, multiple gluon exchange in the $t$-channel gives rise to large high-energy contributions of the type $\alpha_S^k \ln^m 1/x$ (or $\alpha_S^k/N^m$, in the Mellin transform space, $N$ being the moment conjugate to $x$), which may spoil the convergence of the QCD perturbative series at small $x$. These large logarithmic terms have to be identified and summed up to all orders in perturbation the-
ory. One may thus consider an improved perturbative expansion which systematically sums classes of leading $(\mathcal{O}(\alpha_S/N)^k)$, next-to-leading $(\mathcal{O}(\alpha_S(\alpha_S/N)^k))$, etc., $N$-poles (i.e., small-$x$ logarithms) to all orders in $\alpha_S$.

The leading-order resummation can be traced back to the work of Lipatov and collaborators. At leading level, only gluons feed the QCD evolution, and the resummation of the $\mathcal{O}(\alpha_S/N)^k$ contributions to the gluon anomalous dimensions $\gamma_{gg,N}$ is actually accomplished by the BFKL equation [6]. Supplementing the BFKL analysis with the high-energy factorization theorem [7], one can achieve the resummation of the contributions to the coefficient functions as well.

In next-to-leading order, some progress in the evaluation of small-$x$ corrections to both anomalous dimensions and deep inelastic coefficient functions has been obtained recently [8,9], and is reported in the next Section.

## 2. Beyond the leading order

The study of sub-leading corrections at small $x$ is mandatory for estimating and improving the accuracy of the leading-order predictions. In particular, since running-coupling effects mix with small-$x$ sub-leading corrections, it is essential to know the latter in order to be able to carry out consistently the renormalization group analysis in the high-energy regime.

At sub-leading level, quarks start to contribute to the QCD evolution on the same footing as gluons. The current status of resummation is the following. The next-to-leading corrections $\mathcal{O}(\alpha_S(\alpha_S/N)^k)$ to the gluon anomalous dimen-sions are not yet known but a calculational program is underway [10]. The next-to-leading contributions to the quark anomalous dimensions $\gamma_{qg,N}$ have been computed recently [8,9]. The first terms of the perturbative expansion in the $\overline{\text{MS}}$-scheme are ($T_R = 1/2$, $\bar{\alpha}_S = 3\alpha_S/\pi$)

$$\gamma_{qg,N}(\alpha_S) = \frac{\alpha_S}{2\pi} T_R \frac{2}{3} \left\{ 1 + 1.67 \frac{\bar{\alpha}_S}{N} \right. $$
$$+ 1.56 \left(\frac{\bar{\alpha}_S}{N}\right)^2 + 3.42 \left(\frac{\bar{\alpha}_S}{N}\right)^3 + 5.51 \left(\frac{\bar{\alpha}_S}{N}\right)^4 $$
$$\left. + 7.88 \left(\frac{\bar{\alpha}_S}{N}\right)^5 + \mathcal{O}\left(\left(\frac{\bar{\alpha}_S}{N}\right)^6\right) \right\} \ . \quad (1)$$

---

*Contribution at QCD94 Conference, Montpellier, July 1994, presented by F. Hautmann.

Deep inelastic coefficient functions have also been resummed to the next-to-leading accuracy $\mathcal{O}(\alpha_S(\alpha_S/N)^k)$ [9]. The perturbative expansion for the coefficient of $F_2$ in the $\overline{\text{MS}}$-scheme reads (setting the factorization scale $\mu^2$ equal to $Q^2$)

$$C^g_{2,N}(\alpha_S, Q^2/\mu^2 = 1) = \frac{\alpha_S}{2\pi} T_R N_f \frac{2}{3} \left\{ 1 \right.$$
$$+ 1.49 \frac{\bar{\alpha}_S}{N} + 9.71 \left(\frac{\bar{\alpha}_S}{N}\right)^2 + 16.43 \left(\frac{\bar{\alpha}_S}{N}\right)^3$$
$$\left. + 39.11 \left(\frac{\bar{\alpha}_S}{N}\right)^4 + \mathcal{O}\left(\left(\frac{\bar{\alpha}_S}{N}\right)^5\right) \right\} \ . \quad (2)$$

The first two coefficients in Eqs. (1),(2) agree with those known from two-loop calculations [4, 5], whereas the higher-order terms represent new subdominant information at small $x$.

We refer to [9] for a more detailed theoretical discussion of next-to-leading corrections and, in particular, for complete resummed formulae.

A preliminary numerical analysis of next-to-leading effects has recently been performed by the authors of Ref. [11]. They have added the $\mathcal{O}(\alpha_S^3)$ term in Eq. (1) (as well as the first non-vanishing contribution $\mathcal{O}(\alpha_S^4)$ in the BFKL anomalous dimension) to the full two-loop evolution of the parton densities. The resulting predictions for the structure function $F_2$ turn out to depend heavily on the $x$-shape of the input gluon distribution. Whilst the evolution of a steep input (as in the set MRSD_ [12]) is only marginally affected by the inclusion of higher-loop contributions, the evolution of a flat input (as in the set MRSD0 [12]) is in contrast perturbatively unstable, already in the range of $x$-values explored at HERA.

This suggests that further investigations of subleading corrections at small $x$ should be pursued. On one hand, more detailed numerical studies are needed, including the full implementation of resummed results. On the other hand, further theoretical calculations are necessary. In particular, the knowledge of the corrections to the gluon anomalous dimensions is required for performing a fully consistent next-to-leading-order analysis.

**Acknowledgements.** We would like to thank Stephan Narison for the organization of this Conference. This research is supported in part by the EEC Programme *Human Capital and Mobility*, Network *Physics at High Energy Colliders*, contract CHRX-CT93-0357 (DG 12 COMA).